\documentclass[twoside]{article}
\usepackage{fleqn}
 
\def\bvec#1{{\rm\bf #1}}

\usepackage{graphicx}
\usepackage{epsfig}
\usepackage[figuresright]{rotating} 
\begin{document}

\title{Simple solutions of fireball hydrodynamics\\
for self-similar elliptic flows}
\author{
S.V. Akkelin$^{1}$, T. Cs{\"o}rg{\H o}$^2$, B. Luk\'acs$^2$,\\
Yu.M. Sinyukov$^{1}$ and M. Weiner$^3$ }

\date{19.12.2000}
\maketitle

\begin{center}
{\small {\it
 $^1$
Bogolyubov Institute for Theoretical Physics,
Kiev 03143, Metrologicheskaya 14b,Ukraine  \\
 $^2$
MTA KFKI RMKI, H - 1525 Budapest 114, POB 49, Hungary\\
 $^3$
Faculty of Science, E\"otv\"os University,Budapest H-1117, P\'azm\'any P. s. 1/A, Hungary\\
}}
\end{center}

\begin{abstract}
Simple, self-similar, elliptic solutions of non-relativistic
fireball hydrodynamics are presented, generalizing earlier results
for spherically symmetric fireballs with Hubble flows and homogeneous
temperature profiles. The transition from one dimensional to three 
dimensional expansions is investigated in an efficient manner.
\end{abstract}

{\it Introduction.} Recently, a lot of experimental and
theoretical efforts have gone into the exploration of
hydrodynamical behavior of strongly interacting hadronic matter
in non-relativistic as well as in relativistic heavy ion
collisions, see e.g. 
\cite{csernai} - \cite{qm96}. Due
to the non-linear nature of hydrodynamics, exact
hydro solutions are rarely found. Those events,
sometimes, even stimulate an essential progress in physics. One
of the most impressive historical example is Landau's
one-dimensional analytical solution (1953) for relativistic
hydrodynamics \cite{Landau} that gave rise to a new
(hydrodynamical) approach in high energy physics. The boost-invariant 
Bjorken solution \cite{Bjorken}, found more than
20 years later, is frequently utilized as the basis for 
estimations of initial energy densities
in ultra-relativistic nucleus-nucleus collisions.

The obvious success of hydrodynamic approach to high energy
nuclear collisions raise interest in an analogous
description of non-relativistic collisions, too. 
The first exact non-relativistic hydrodynamic solution describing
expanding fireballs was found in 1979 \cite{jozso}. 
It has been generalized for fireballs with 
Gaussian density and homogeneous temperature 
profiles~\cite{elozo cikk} as well as for fireballs with
arbitrary initial temperature profiles~\cite{solrt} and corresponding,
non-Gaussian density profiles.  
All of these solutions have spherical symmetry and a Hubble-type
linear radial flow.
However, a non-central collision has none of the mentioned
symmetries.
The purpose of this Letter is to present and analyze
hydro solutions for such cases. The results
presented in this paper may be utilized to access 
the time-evolution of the hydrodynamically behaving, strongly
interacting matter as probed by non-central non-relativistic heavy ion
collisions \cite{nr,nrt}. As the hydro equations have no
intrinsic scale, the results are rather
general in nature and can be applied to any physical
phenomena where the non-relativistic hydrodynamical description
is valid. 

{\it Generalization of spherical solutions to elliptic flows.}
Consider an ideal fluid, 
where viscosity and heat conductivity are negligible, described
by the local conservation of matter ({\it continuity equation}), the
local conservation of energy ({\it energy equation}) and the local
conservation of momentum ({\it Euler equation}):
\begin{eqnarray}
{\frac{\partial n}{\partial t}}+\bvec \nabla (\bvec%
v n) &=&0,  \label{continuity-eq} \\
{\frac{\partial \varepsilon }{\partial t}}+\bvec \nabla (%
\bvec v \varepsilon ) &=&-P\bvec \nabla %
\bvec v,  \label{energy-eq} \\
mn({\frac \partial {\partial t}}+\bvec v \bvec%
\nabla)\bvec v &=&-{\bvec%
\nabla P},  \label{Euler-eq}
\end{eqnarray}
where $m$ is the mass of a single particle and $n=n(t,\bvec r%
),$ $\bvec v =\bvec v(t,\bvec%
r),$ $P=P(t,\bvec r)$ and $\varepsilon
=\varepsilon (t,\bvec r)$ are the \textit{local} density of
the particle number, the velocity, the pressure and the energy density
fields, respectively. To complete the set of equations
we need to fix the equations of state. For reasons of simplicity we have
chosen the equations of state to be those of an ideal, structureless Boltzmann
gas:
\begin{eqnarray}
\varepsilon (t,\bvec r) &=&{\frac 32}P(t,\bvec%
r),  \label{energy-density} \\
P(t,\bvec r) &=&n(t,\bvec r)T(t,\bvec%
r).  \label{EOS}
\end{eqnarray}
The solutions that are presented in the subsequent parts cannot be
trivially generalized to any arbitrary equations of state. Nevertheless,
they provide a transparent insight into the collective physical processes
in a non-central heavy ion collision.

In the following, we utilize the above ideal gas equation 
of state and rewrite the hydrodynamical equations in terms of 
the three functions $n$,$\bvec v$ and $T$.

In ref. \cite{elozo cikk,solrt}, special classes of exact analytic
solutions of fireball hydrodynamics were found
assuming spherical symmetry and self-similar Hubble flows.
In ref.~\cite{elozo cikk} a homogeneous temperature profile was assumed,
while the general solution for arbitrary, inhomogeneous 
initial temperature profiles was found in ref.~\cite{solrt}. 
In these articles, the concept of self-similarity
meant that there is a typical length-scale of the expanding system
$R=R(t)$ so that all space-time functions in the hydro equations are of
the form $F=G(t)H(s)$ where $s=r^2/R^2$ is the so-called
(dimensionless) scaling variable.

Let us go beyond spherical symmetry and consider three
typical lengths of the expanding system: $X,Y$ and $Z$, all
functions of time only.  Let us rotate our frame of reference to
the major axis of the ellipsoidal expansion, and leave to 
future applications to relate these major axes to the laboratory frame. 
Consequently, let us  
introduce three scaling variables $x=r_x^2/X^2$, $y=r_y^2/Y^2$ and $%
z=r_z^2/Z^2$ and assume that all space-time functions are of the form of 
$ F=G(t)H(x)K(y)L(z)$. Using this ansatz we find that the \textit{continuity
equation} is satisfied regardless of the density profile if the velocity
field is a Hubble-flow field in each principal direction:
\begin{eqnarray}
v_x(t,\bvec r) &=&{\frac{\dot{X}(t)}{X(t)}}r_x,  \nonumber
\label{v1} \\
v_y(t,\bvec r) &=&{\frac{\dot{Y}(t)}{Y(t)}}r_y,  \nonumber
\label{v2} \\
v_z(t,\bvec r) &=&{\frac{\dot{Z}(t)}{Z(t)}}r_z.  \label{v}
\end{eqnarray}
Although our sole assumption concerning the temperature was the ansatz form
already mentioned, we found that the \textit{Euler equation} requires the
temperature to be homogeneous, independent of the coordinate variables: $%
T=T(t)$. The energy equation is only satisfied if
\begin{eqnarray}
T(t)=T_0\left( {\frac{\displaystyle\phantom{|}V_0}{\displaystyle\phantom{|}%
V(t)}}\right) ^{2/3}  \label{T},
\end{eqnarray}
where $V(t)=X(t)Y(t)Z(t)$ is the typical volume of the expanding system,
while $V_0=V(t_0)$ and $T_0=T(t_0)$ are the initial temperature and volume.
The homogeneity of the temperature and the Euler equation implied that the
density profile is a product of three Gaussians, with different, time
dependent radius parameters:
\begin{eqnarray}
n(t,\bvec r)=n_0{\frac{\displaystyle\phantom{|}V_0}{%
\displaystyle\phantom{|}V(t)}}\exp \left( -{\frac{\displaystyle\phantom{|}%
r_x^2}{\displaystyle\phantom{|}2X(t)^2}}-{\frac{\displaystyle\phantom{|}r_y^2%
}{\displaystyle\phantom{|}2Y(t)^2}}-{\frac{\displaystyle\phantom{|}r_z^2}{%
\displaystyle\phantom{|}2Z(t)^2}}\right),   \label{n}
\end{eqnarray}
where $n_0=n(0,\bvec 0)$ can be expressed by the total number of
particles ($N$) as
\begin{eqnarray}
n_0=\frac{1}{(2\pi )^{3/2}}\frac{N}{V_0}.  \label{n0}
\end{eqnarray}
The time evolution of the scales are determined - through the \textit{Euler
equation} - by the equations
\begin{equation}
\ddot{X}X=\ddot{Y}Y=\ddot{Z}Z=\frac{T_0}m\left( {\frac{\displaystyle%
\phantom{|}V_0}{\displaystyle\phantom{|}V}}\right) ^{2/3} . \label{scales}
\end{equation}
This system of non-linear, second-order ordinary differential equations 
has a unique solution for the scale-functions 
if the initial parameters $X_0$, $Y_0$, $Z_0
$ and $\dot{X}_0$, $\dot{Y}_0$, $\dot{Z}_0$ are given. 
Although this solution 
has not yet been found in an explicite, analytic form, 
some of its properties are  determined in the subsequent parts.

%
{\it Properties of the elliptic solutions. } 
Global conservation laws  reflect, in general, boundary conditions for
solutions, or their behavior at asymptotically large distances.
Because of reflection symmetry of densities and velocities, 
the conservation of momentum, $\bvec P=0$, 
is satisfied automatically and gives no
non-trivial first integral. On the other hand, the
asymptotically fast decreasing of the densities give us the
possibility of using total energy conservation as

\begin{equation}
\frac \partial {\partial t}\int d^3r(\varepsilon +\frac{nmv^2}2)=0
\label{energy}
\end{equation}
to find the first integral of the system of equations (\ref{continuity-eq})
- (\ref{Euler-eq}). Substitute (\ref{v}), (\ref{n}) into (\ref{energy}), we
get:
\begin{equation}
\stackrel{.}{X}^2+\stackrel{.}{Y}^2+\stackrel{.}{Z}^2+3\frac{T_0}m\left( {%
\frac{\displaystyle\phantom{|}V_0}{\displaystyle\phantom{|}V}}\right)
^{2/3}=A\, =const.  \label{relation}
\end{equation}
Using (\ref{scales}) one can rewrite (\ref{relation}) in the form
\begin{equation}
\frac 12\frac{\partial ^2}{\partial t^2}(X^2+Y^2+Z^2)=A,  \label{relation1}
\end{equation}
and find finally
\begin{equation}
R^2(t) = X^2(t)+Y^2(t)+Z^2(t) \, =\, A(t-t_0)^2+B(t-t_0)+C,  \label{first integ}
\end{equation}
where
\begin{eqnarray}
A &=&\stackrel{.}{X}_0^2+\stackrel{.}{Y}_0^2+\stackrel{.}{Z}_0^2+3\frac{T_0}%
m,  \nonumber \\
B &=&2(X_0\stackrel{.}{X}_0+Y_0\stackrel{.}{Y}_0+Z_0\stackrel{.}{Z}_0),
\nonumber \\
C &=&X_0^2+Y_0^2+Z_0^2.  \label{contants}
\end{eqnarray}

The simple equation (\ref{first integ}) express the general
property of elliptic hydrodynamic flows. The value of
radius-vector evolves in time similar to a ''particle'' with
coordinates $(X,Y,Z)$ that moves in a non-central, repulsive 
potential according to
eqs. (\ref{scales}). In particular, one may introduce the canonical
coordinates $(X,Y,Z)$ and the canonical momenta as $(P_x, P_y, P_z) = m 
(\dot{X}, \dot{Y}, \dot{Z})$ and the Hamiltonian $ H $ as a
rewritten form of eq. ~(\ref{relation}):
\begin{equation} 
H = \frac{1}{2m} \left( P_x^2 + P_y^2 + P_z^2\right) + \frac{3}{2} T_0
\left( \frac{X_0 Y_0 Z_0}{X Y Z} \right)^{2/3}
\end{equation} 
The Hamiltonian equations of motion can be written in terms of Poisson
brackets as $\dot{X} = \left\{X,H\right\}$, ... , 
$\dot{P_x} = \left\{ P_x, H\right\}$, ... . The Lagrangian form of
these equations is given by eqs.~(\ref{scales}). Due to the repulsive 
nature of the potential, the coordinates  $(X,Y,Z)$ diverge to infinity
for large times. As the potential vanishes for large values of the
coordinates, the canonical momenta tend to constant values for asymptotically
large times.

Eq. (\ref{relation}) expresses the
conservation of whole energy (kinetic and potential) of the
''particle'', corresponding to $H(X, ..., P_x, ... ) = E = const$. 
The resulting eq.~(\ref{first integ})  has also  great
importance for the  analysis of approximate analytical solutions. It
is worth mentioning another interesting relation that one can
get from (\ref{relation}) for asymptotic times $t_{as},$ when
$V(t_{as})\gg V_0$:
\begin{equation}
\frac{m}{2}\left(\stackrel{.}{X}_{as}^2
+\stackrel{.}{Y}_{as}^2+\stackrel{.}{Z}_{as}^2\right) =
\frac{m}{2}\left(\stackrel{.}{X}_0^2
+\stackrel{.}{Y}_0^2+\stackrel{.}{Z}_0^2\right)
+\frac{3}{2}T_0.
\label{relation-vel}
\end{equation}
This relation expresses the equality of the initial flow and internal
energy with the asymptotic energy which is present in the form of flow.

Although eqs.~(\ref{scales}) are easy to handle with presently available
numerical packages,  we note that a further simplification of these 
equations to a non-linear first order differential equation of one variable
is possible, if an additional cylindrical symmetry is assumed, corresponding to 
$X(t) = Y(t)$.
One may introduce the angular variable $\phi = \arccos(Z/R)$ so that
\begin{eqnarray}
	X(t) & = & Y(t) \, = \,\frac{1}{\sqrt{2}} R(t) \sin \phi(t) \\
	Z(t) & = & R(t) \cos\phi(t).
\end{eqnarray}
The time
evolution of $\phi(t) $ is determined by the following first order equation:
\begin{equation}
	\dot{\phi}^2 = \frac{1}{R^2(t)} \left[\frac{E}{m} - \dot{R}^2(t) -
	\frac{3}{2} \frac{T_0}{m} \frac{(X_0^2 Z_0)^{2/3} }{R^2(t)}
	\frac{1}{(\sin\phi)^{4/3}\, (\cos\phi)^{2/3} } \right]
\end{equation} 
where $R(t)$ is given explicitly by eq.~(\ref{first integ}).

\begin{figure}
\vspace{-1.5truecm}
\begin{center}
\epsfig{file=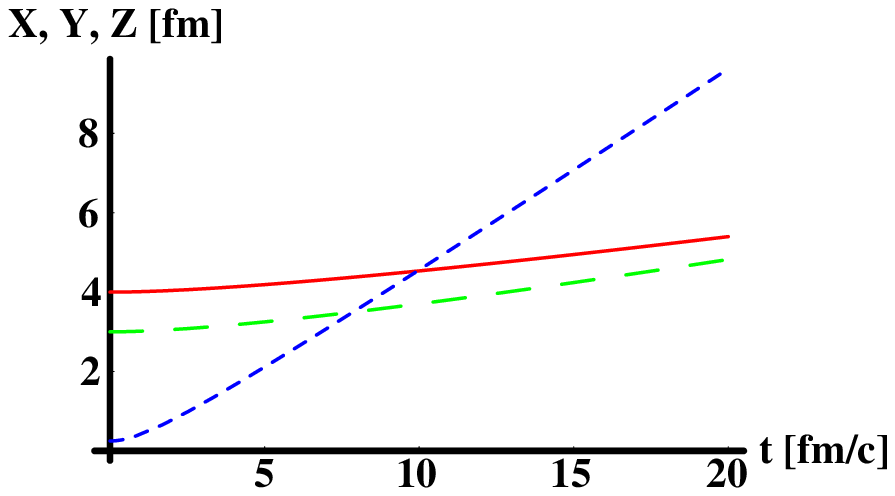,width=4.5in,angle=0}
\epsfig{file=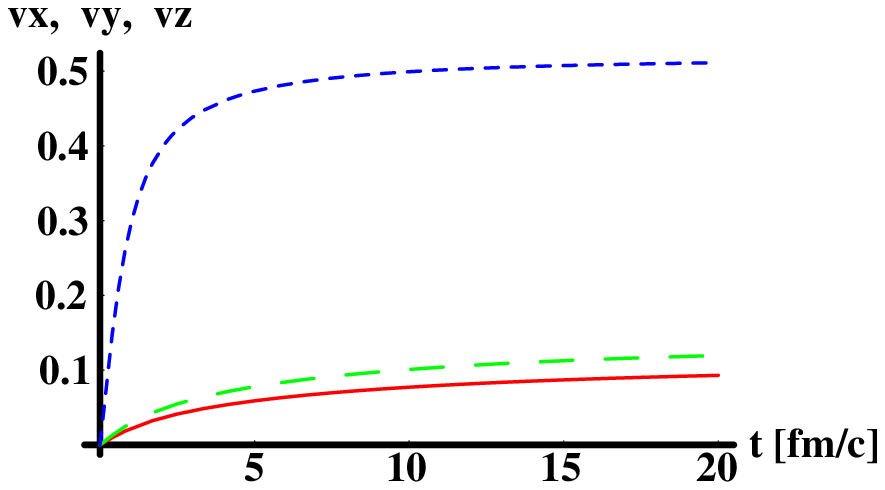,width=4.5in,angle=0}
\end{center}
\caption{\label{f:1} Top panel shows the time dependence of the major axis of the ellipsoidal
(Gaussian) density profile, eq.~(\ref{n}),
while the bottom panel indicates the velocity of expansion at the rms radii,
$r_x = X(t)$, $r_y = Y(t)$ and $r_z = Z(t)$, 
 for three dimensional, self-similar elliptic flows.
Solid lines stand for $X(t)$ and $v_x(t)$, dashed lines for $Y(t)$
and $v_y(t)$, short-dashed lines for
$Z(t)$ and $v_z(t)$.  
	The initial conditions are: $X_0 = 4$ fm, $Y_0 = 3$ fm, 
$Z_0 = 1/4$ fm, the initial velocities are all vanishing, $T_0/m = 0.1$ .
Due to the Landau-like initial condition (strong initial compression in the 
$z$ direction), the flow is almost one dimensional, only small amount of
transverse flow is generated in this case.}
\end{figure}

\begin{figure}
\vspace{-1.5truecm}
\begin{center}
\epsfig{file=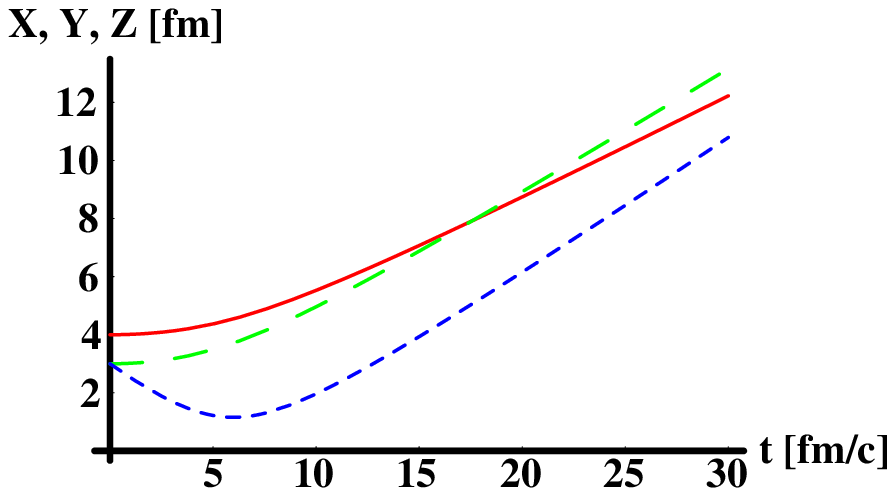,width=4.5in,angle=0}
\epsfig{file=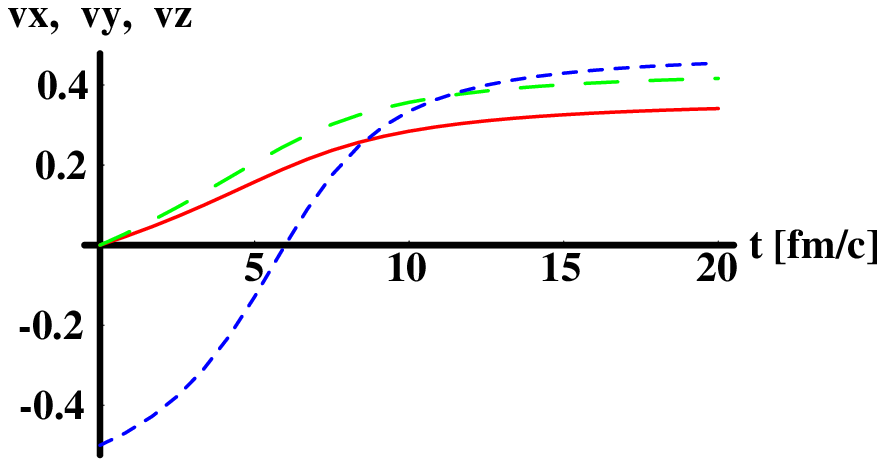,width=4.5in,angle=0}
\end{center}
\caption{\label{f:2} 
Same as Fig.~\ref{f:1}, but for a more spherical initial
profile with an initial inwards flow in the $z$ direction.
The initial conditions are: $X_0 = 4$ fm, $Y_0 = 3$ fm, 
$Z_0 = 3$ fm, the initial velocities are $\dot X_0 =0$, $\dot Y_0 = 0$,
$\dot Z_0 = -0.5$, while $T_0/m = 0.1$ .
Due to the deviation from the Landau-like initial condition, 
the final flow is almost spherically symmetric, three dimensional, 
and a large amount of transverse flow is generated in this case.}
\end{figure}

\begin{figure}
\vspace{-1.5truecm}
\begin{center}
\epsfig{file=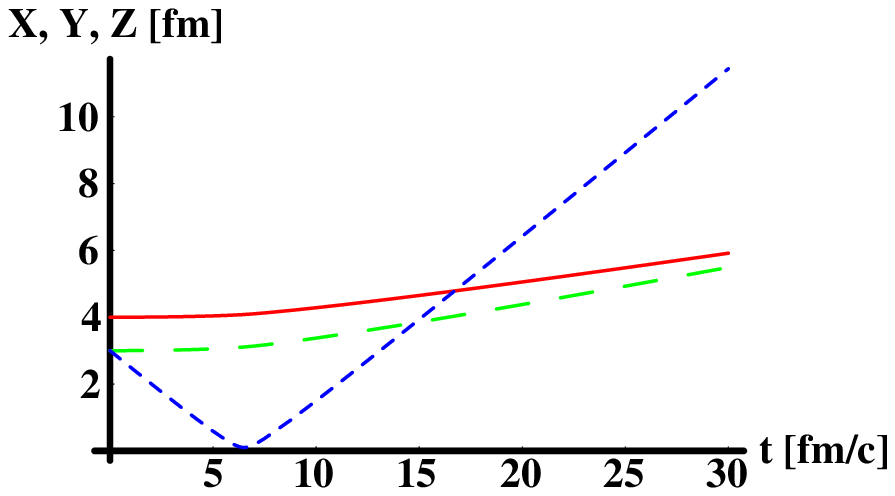,width=4.5in,angle=0}
\epsfig{file=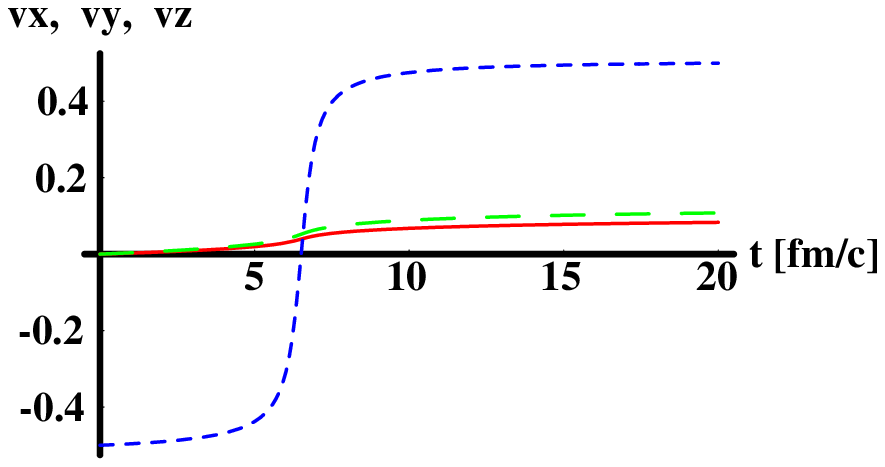,width=4.5in,angle=0}
\end{center}
\caption{\label{f:3} 
Same as Fig.~\ref{f:3}, but for $T_0/m = 0.01$.
This corresponds to increasing the initial kinetic energy
as compared to the internal (thermalized) energy,
and in this case the flow becomes approximately one dimensional again.
The initial conditions are: $X_0 = 4$ fm, $Y_0 = 3$ fm, 
$Z_0 = 3$ fm, the initial velocities are $\dot X_0 =0$, $\dot Y_0 = 0$,
$\dot Z_0 = -0.5$, while $T_0/m = 0.01$ .}
\end{figure}

\begin{figure}
\vspace{-1.5truecm}
\begin{center}
\epsfig{file=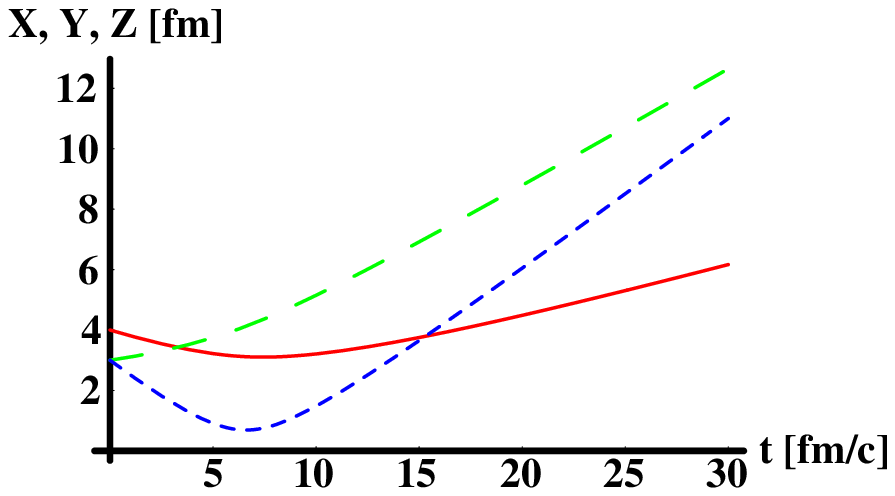,width=4.5in,angle=0}
\epsfig{file=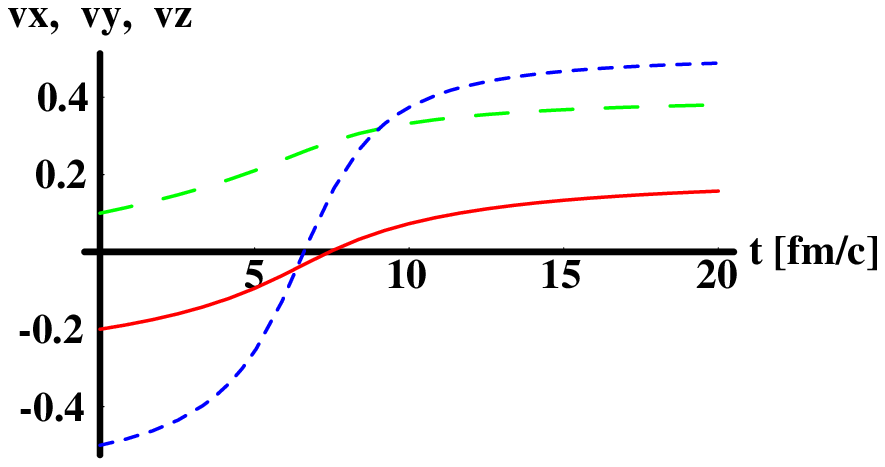,width=4.5in,angle=0}
\end{center}
\caption{\label{f:4} 
A more general initial condition may result even in a dominantly 
two-dimensional expansion.
The initial conditions are: $X_0 = 4$ fm, $Y_0 = 3$ fm, 
$Z(0) = 3$ fm, the initial velocities are $\dot X_0 = -0.2$, $\dot Y_0 = 0.1$,
$\dot Z_0 = -0.5$, while $T_0/m = 0.05$ .}
\end{figure}

Figs.~\ref{f:1}-\ref{f:4} indicate the results of numerical solutions of
eqs.~(\ref{scales}). Using Landau-type initial conditions, one
confirms that even the full three-dimensional solution results in
a small amount of  transverse flow generation, while for more general
initial conditions, significant amount of transverse flow can be generated.
Transverse flow is stronger if the initial conditions are closer
to spherical symmetry or, if the  fraction of the initial thermal energy
is increased as compared to the initial kinetic energy. For more details,
see the figure captions.

In the last part an approximate,  analytic solution is presented that
corresponds  to Landau-like, one dimensional expansions.

{\it Approximate one-dimensional solutions. }
Hydrodynamical evolution, which is described
by Eqs. (\ref{continuity-eq}) - (\ref{Euler-eq}), starts from some initial
conditions. Consider Landau-type initial conditions (longitudinally
compressed ''ellipsoid''), corresponding to the real situation in
non-relativistic heavy ion collisions:
\begin{equation}
\stackrel{.}{X}_0=\stackrel{.}{Y}_0=0,\, Z_0\ll X_0,\,Z_0\ll Y_0,
\label{initial}
\end{equation}
and in general case $\stackrel{.}{Z}_0\neq 0$. The last reflects the
situation when a system is (locally) thermalized 
before, after or at the
moment of full nuclear stopping and transverse expansion 
starts to develop only after the local thermalization. 
Then in some time interval $t_0\leq t\leq \widetilde{t}$ the hydro evolution 
is quasi - one - dimensional:
\begin{equation}
\frac{X(t)}{X_0}\approx 1,\,\frac{Y(t)}{Y_0}\approx 1,\, 
\label{1-dm}
\end{equation}
and the  equation of motion for $Z(t)$ takes the following form
\begin{equation}
\ddot{Z}Z=\frac{T_0}m\left( {\frac{\displaystyle\phantom{|}Z_0}{\displaystyle%
\phantom{|}Z}}\right) ^{2/3}.  \label{1dm-eq}
\end{equation}
This equation has an exact analytic solution,
\begin{equation}
Z^2(t)=(\widetilde{Z}_0^{2/3}+(a_{+}+a_{-})^2)^3,  \label{solution}
\end{equation}
where
\begin{eqnarray}
a_{\pm } &=&\left( \frac 12\widetilde{Z}(t)\pm \left( \widetilde{Z}%
_0^2+\frac 14\widetilde{Z}^2(t)\right) ^{1/2}\right) ^{1/3},  \nonumber \\
\widetilde{Z}(t) &=&\sqrt{3u^2}(t-\widetilde{t}_0),\,u^2=\frac
13\left( \stackrel{.}{Z}_0^2+3\frac{T_0}m\right) ,  \nonumber \\
\widetilde{Z}_0 &=&\left( \frac{T_0}{mu^2}\right) ^{3/2}Z_0,\,%
\widetilde{t}_0=t_0-\frac{Z_0\stackrel{.}{Z}_0}{3u^4}(u^2+\frac{T_0}m).
\label{1dm-param}
\end{eqnarray}
Here $\widetilde{t}_0$ is 
the turning point for $Z^2(t)$ if $\stackrel{.}{Z}_0<0  $. 
Using (\ref{first integ}) we obtain that 
the conditions for validity of the solution (\ref{1-dm}), (\ref
{solution}) are  satisfied within some time interval $t_0\leq
t\leq \widetilde{t}$ if
\begin{equation}
\left| \frac{A(t-t_0)^2+B(t-t_0) + Z_0^2 - Z^2(t)}{X_0^2+Y_0^2}\right| \ll 1.
\label{1dm-condition}
\end{equation}

The hydrodynamical evolution described by the equations (\ref{continuity-eq}%
) - (\ref{Euler-eq}) cannot be continued infinitely in time, because
the general criteria of applicability of hydrodynamical
description are violated: the mean free path $l\propto 1/(\sigma n)$ has to
be (much) smaller than the typical length scales of the system, for example
the effective geometrical sizes or hydrodynamic lengths. 
Due to the hydrodynamical expansion the density (\ref{n}) will
decrease with time reaching some critical value
that can be estimated utilizing
Landau's criterium, $T=m_\pi ,$ that determines a critical
density when hydrodynamic evolution breaks up.
Here, we will use a
simplified version of this criterium 
and suppose the decoupling of hydrodynamical
system when density in the center of the system reaches some
critical value $n_f$ (typically, normal nuclear density). Then
the time $t_{f\,}$, when the hydrodynamical evolution ends,
can be estimated from the condition
\begin{equation}
n(t_f,\bvec 0)=n_f,  \label{freeze-out}
\end{equation}
If the hydrodynamical evolution stops before the condition 
(\ref{1dm-condition}) is
violated, $t_{f\,}<\widetilde{t}$, then solutions (\ref{1-dm}) and (%
\ref{solution}) describing quasi-one-dimensional
expansion give complete hydrodynamical evolution, too. 
Let us find the conditions for such a situation. 
Supposing quasi-one-dimensional expansion and using
(\ref{n}) and (\ref{1-dm}) we get
\begin{equation}
\frac{Z(t_f)}{Z_0}=\frac{n_0}{n_f}.  \label{freeze-out-cond}
\end{equation}
Then using (\ref{solution}) we can find $t_f$. If for $t=t_f$ the inequality
(\ref{1dm-condition}) is satisfied, then we can conclude that $t_{f\,}<%
\widetilde{t}$ and the one dimensional expansion is valid approximation
until the freeze-out time.

It is useful to give simple analytical estimations of the initial
hydrodynamic conditions that guarantee quasi-one-dimensional expansion of
the nuclear matter. Let us suppose, for simplicity, that $\stackrel{.}{Z}_0=0
$ and hence $u^2=\frac{T_0}m,$ $\widetilde{Z}_0=Z_0$, $\widetilde{t}_0=t_0$
in (\ref{solution}). Supposing that the upper time limit of
quasi-one-dimensional expansion, $\widetilde{t},$ is large enough so that
$$
\widetilde{Z}^2(\widetilde{t})\gg 4Z_0^2,  
$$
we can rewrite Eq.(\ref{solution}) in the form:
\begin{equation}
Z^2(\widetilde{t})\approx \widetilde{Z}^2(\widetilde{t})-3(Z_0\widetilde{Z}%
^2(\widetilde{t}))^{2/3}.  \label{z-approx}
\end{equation}
After substitution (\ref{z-approx}) in (\ref{1dm-condition}) we obtain:
\begin{equation}
\frac{3\frac{T_0}m(\widetilde{t}-\widetilde{t}_0)^2}{X_0^2+Y_0^2}\ll \frac
1{3\sqrt{3}}\frac{\sqrt{X_0^2+Y_0^2}}{Z_0}.  \label{cond-approx2}
\end{equation}
It is easy to see from (\ref{cond-approx2}) that for large enough $\frac{%
X_0^2+Y_0^2}{Z_0^2}$ , quasi-one-dimensionality can hold till the 
longitudinal scale becomes comparable to transversal ones:
\begin{equation}
\frac{Z^2(\widetilde{t})}{X^2(\widetilde{t})+Y^2(\widetilde{t})}\approx
\frac{\widetilde{Z}^2(\widetilde{t})}{X_0^2+Y_0^2}\approx \frac{Z^2(%
\widetilde{t})}{X_0^2+Y_0^2}\propto 1  \label{final}
\end{equation}
It means that within time $t<$ $\widetilde{t}$ solution (\ref{1-dm}), (\ref
{solution}) correctly describes the transformation of longitudinally
compressed ''ellipsoid'' to ''spheroid'' form. Finally
from (\ref{freeze-out-cond}) we get that $Z(t_f)<Z(\widetilde{t})\propto
\sqrt{X_0^2+Y_0^2}$ and consequently $t_f<\widetilde{t}$ if
\begin{equation}
\frac{n_0}{n_f}<\frac{\sqrt{X_0^2+Y_0^2}}{Z_0}  \label{density-cond}
\end{equation}

Under such initial conditions one can expect that whole stage of the
hydrodynamical evolution can be correctly described by the approximate
quasi-one-dimensional solution (\ref{1-dm}) and (\ref{solution}).

{\it Summary and conclusions:}
	In this Letter we considered the time evolution of  fireball  
	hydrodynamics describing an ideal gas, an elliptic initial density
	profile, a homogeneous temperature distribution and a Hubble-like
	flow distribution. For this case, the set of partial differential
	equations of non-relativistic hydrodynamics have been reduced 
	to a set of ordinary, second order, non-linear differential equations,
	that can be solved efficiently by presently available numerical
	packages without the need of sophisticated programming.
	The initial conditions for these equations are associated with the
	initial elliptic sizes $X_0$, $Y_0$ and $Z_0$, that could be linked
	with the overlapping geometrical sizes of colliding nuclei
	in heavy ion collisions and with the dynamics of the compression
	process during interactions during the pre-thermal time evolution. 

	The general behavior of these hydrodynamical equations is determined
	analytically and related to the Hamiltonian motion of a particle
	in a repulsive, non-central potential. A first integral of motion 
	has been found, corresponding to the conservation of energy
	in the Hamiltonian problem.  It was utilized to obtain 
	an approximate solution for quasi one-dimensional 
	expansions and to determine the domain of applicability
	of this solution.

	The importance of the results is given by recent experimental
	findings in high energy heavy ion reacions, where various 
	elliptic flow patters are observed, see 
	refs.~\cite{poskanzer,ack,csernai-99,voloshin} for further 
	details.
	In future studies, our results could be 
	applied to gain insight into the interpretation
	of the above mentioned data and to describe nucleus-nucleus collisions
	with non-relativistic initial energies.
 	Such conditions
	for a non-relativistic evolution may be reached in the mid-rapidity
	region near to the softest point of equation of state even in
	relativistic heavy ion collisions, if the 
	pressure is not strong enough to build up a relativistic transverse
	 flow. Due to the scale invariance of the hydrodynamical equations
        the solutions described here can also be utilized in other problems
	related to elliptic flows in non-spherical fireball hydrodynamics.
	
{\it Acknowledgments:}
        This research has been supported by a Bolyai Fellowship
        of the Hungarian Academy of Sciences and by the grants OTKA
        T024094, T026435, T029158, the US-Hungarian Joint
        Fund MAKA grant 652/1998, NWO - OTKA
        N025186, Hungarian - Ukrainian  S\& T grant 45014 (2M/125-199)
        and the grants FAPESP 98/2249-4 and 99/09113-3 of Sao Paolo, Brazil.                                   

\end{document}